\documentclass[letterpaper ,10pt, conference]{IEEEtran}      

\IEEEoverridecommandlockouts                              

\date{}

\newcommand{\argmax}{\operatornamewithlimits{argmax}}

\usepackage{graphics} 
\usepackage{epsfig} 
\usepackage{amsmath} 
\usepackage{amssymb}  
\usepackage{amssymb,amsmath}
\usepackage{graphicx,color}

\newcommand{\x}{{\bf x}}

\newcommand{\z}{{\bf z}}
\newcommand{\q}{{\bf q}}
\newcommand{\Ex}{{\mathbb{E}}}

\newcommand{\prob}{{\mathsf P}}
\newcommand{\X}{{\bf {X}}}
\newcommand{\Z}{{\bf {Z}}}
\newcommand{\N}{{\bf {N}}}

\newcommand{\tends}{\rightarrow}

\newcommand{\qed}{\hspace*{\fill}~{\rule{2mm}{2mm}}\par\endtrivlist\unskip}

\newcommand{\complex}{{\mathbb C}}

\newcommand{\snr}{{\mathsf{SNR}}}

\begin{document}

\title{ On Block Noncoherent Communication with Low-Precision Phase Quantization at the Receiver \vspace{-2mm}}

\author{Jaspreet Singh and Upamanyu Madhow \vspace{-2mm}$^*$
\thanks{}
\thanks{$^*$This work was supported in part by the National Science Foundation under grants CCF-0729222 and CNS-0832154, and by the Office for Naval Research under grant N00014-06-1-0066. The authors are with the ECE Department, UC Santa Barbara, CA 93106, USA.
{\tt\small  \{jsingh, madhow\}@ece.ucsb.edu}.}
}

\maketitle

\begin{abstract} We consider communication over the block noncoherent
AWGN channel with low-precision Analog-to-Digital Converters (ADCs) at the receiver. For standard uniform Phase Shift Keying (PSK) modulation, we investigate the performance of a receiver architecture that quantizes {\it only the phase} of the received signal; this has the advantage of being {\it implementable without automatic gain control}, using multiple 1-bit ADCs preceded by analog multipliers. We study the structure of the transition density of the resulting channel model. Several results, based on the symmetry inherent in the channel, are provided to characterize this transition density. A low complexity procedure for computing the channel capacity is obtained using these results. Numerical capacity computations for QPSK show that $8$-bin phase quantization of the received signal recovers more than $80$-$85$\% of the capacity attained with unquantized observations, while $12$-bin phase quantization recovers above $90$-$95$\% of the unquantized capacity. Dithering the constellation is shown to improve the performance in the face of drastic quantization.
\end{abstract}

\vspace{-1mm}
\section{Introduction}
As communication systems scale up in speed and bandwidth, the cost and
power consumption of high-precision Analog-to-Digital Conversion (ADC)
becomes the limiting factor in modern receiver architectures based on
Digital Signal Processing (DSP) \cite{Walden}. One possible approach
for the design of such DSP-centric architectures is to reduce the
precision of the ADC. In our prior work \cite{ISIT08, TCOM_draft}, we
analyzed the impact of low-precision quantization on the capacity of
the ideal real baseband discrete-time Additive White Gaussian Noise
(AWGN) channel. In this paper, we consider a {\it block noncoherent}
complex baseband AWGN channel that models the effect of carrier
asynchronism. If the receiver's local oscillator is not synchronized
with that of the transmitter, the phase after downconversion is a
priori unknown, but, for practical values of carrier offset, well
approximated as constant over a block of symbols. \looseness-1

The classical approach to noncoherent communication is to approximate
the phase as constant over two symbols, and to apply differential
modulation and demodulation. Divsalar and Simon \cite{Divaslar} were the first to point out the gains that may be achieved by performing multiple symbol
differential demodulation over a block of $L>2$ symbols. More recent work \cite{Warrier, Sweldens, Rong-Rong} has shown that block demodulation, even for large values of $L$, can be implemented efficiently, and exhibits excellent performance for both coded and uncoded systems.

In this work, we study the effect of low-precision receiver quantization for the block noncoherent AWGN channel, under $M$-ary Phase Shift Keying (MPSK) modulation. Since PSK encodes the information in the phase of the transmitted symbol, we investigate an architecture in which the receiver simply quantizes the phase of the received signal, disregarding the amplitude information. Such phase quantization can be efficiently implemented using $1$-bit ADCs preceded by analog multipliers: the use of $1$-bit ADCs is attractive since it results in significant power savings and also eliminates the need for Automatic Gain Control (AGC). We study the structure of the input-output relationship of the resulting {\it phase quantized}-block noncoherent AWGN channel. Based on the
symmetry inherent in the channel model, we derive several results characterizing the output probability distribution over a block of symbols, both conditioned on the input, and without conditioning. These results are used to provide a low-complexity procedure for computing the capacity of the channel (brute force computation has complexity exponential in block length $L$). As in prior work on the unquantized block noncoherent channel, our capacity computations assume that the channel phase is independent from block to block, (this yields a pessimistic estimate of performance, since the phase correlation across blocks can, in principle, be exploited to improve performance).  Numerical results are provided for Quaternary Phase Shift Keying (QPSK) with 8-bin and 12-bin phase quantization at the receiver, and compared with the unquantized capacity obtained earlier in \cite{Peleg}. We also provide results that indicate that dithering the constellation improves performance in the face of drastic quantization. \looseness-1

{\it Notation}: Throughout the paper, we denote random variables by
capital letters, and the specific value they take using small
letters. Bold faced notation is used to denote vectors of random
variables. $\Ex$ is the expectation operator.

\section{Channel Model and Receiver Architecture}

The received signal over a block of length $L$, after quantization is represented as
\begin{equation}\label{eq:Channel_Model}
{Z_l} = \mathsf{Q}(S_le^{j\Phi} +N_l) \ , \ l=0,1,\cdots,L-1,
\end{equation}
where,
\begin{itemize}
\item ${\bf S}:=[S_0  \ S_1  \ \cdots  \ S_{L-1}]$ is the transmitted vector,
\item $\Phi$ is an unknown constant with uniform distribution on $[0, 2\pi)$,
\item $\N:=[N_0 \  \cdots \  N_{L-1}]$ is a vector of i.i.d. complex Gaussian noise with variance $\sigma^2=N_0/2$ in each dimension,
\item $\mathsf{Q}:\complex \tends \mathcal{K}=\{0,1,\cdots,K-1\}$ denotes a quantization function that maps each point in the complex plane to one of the $K$ quantization indices, and
\item $\Z:=[Z_0 \  Z_1 \  \cdots \  Z_{L-1}]$ is the vector of quantized received symbols, so that each $Z_l \in \cal{K}$.
\end{itemize}
Each $S_l$ is picked in an i.i.d. manner from a uniform M-PSK
constellation denoted by the set of points
$\mathcal{A}=\{e^{j\theta_0}, e^{j\theta_1}, \cdots,
e^{j\theta_{M-1}}\}$, where $\theta_m=(\theta_{m-1}+\frac{2\pi}{M})$
{\footnote{Unless stated otherwise, any arithmetic operations for
phase angles are assumed to be performed modulo $2\pi$. For the output
symbols $Z_l$, the arithmetic is modulo $K$, while for the input
symbols $X_l$ (introduced immediately after in the text ), it is
modulo M.}}, for $m=1,2,\cdots,M-1$.

We now introduce the random vector $\X=[X_0 \ \ X_1 \ \ \cdots \ \ X_{L-1}]$, with each $X_i$ picked in an i.i.d. manner from a uniform distribution on the set $\{0,1,\cdots,M-1\}$. Our channel model \eqref{eq:Channel_Model} can now equivalently be written as
\begin{equation}\label{eq:Channel_Model_e}
{Z_l} = \mathsf{Q}(e^{j\theta_{X_l}}e^{j\Phi} +N_l) \ , \ l=0,1,\cdots,L-1 \ ,
\end{equation}
with every output symbol $Z_l \in \{0,1,\cdots,K-1\}$ as before, and
every input symbol $X_l \in \{0,1,\cdots,M-1\}$. The set of all
possible input vectors is denoted by $\mathcal{X}$, while
$\mathcal{Z}$ denotes the set of all possible output vectors.

\begin{figure}[t]
      \centerline { \epsfig{figure=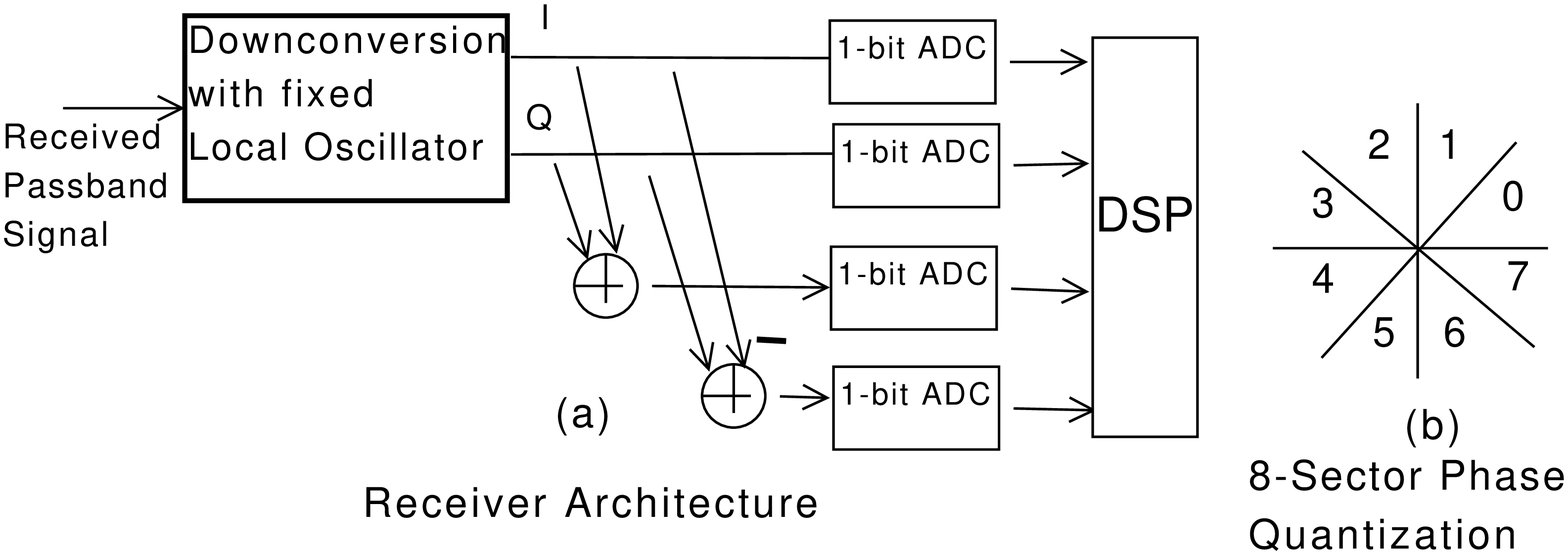, width=9cm}}
      \vspace{-5mm}\caption{\footnotesize Receiver architecture for 8-sector quantization.}
      \vspace{-5mm}
       \label{fig:Receiver}
 \end{figure}

We consider $K$-bin (or $K$-sector) phase quantization: our quantizer
divides the interval $[0,2\pi)$ into $K$ equal parts, and the
quantization indices go from $0$ to $K-1$ in the counter-clockwise
direction. Fig. \ref{fig:Receiver}(b) depicts the scenario for $K$=8.
Thus, our quantization function is $\mathsf{Q}(c)=\lfloor
\arg(c)|(\frac{2\pi}{K}) \rfloor$, where $c \in \complex$, and
$\lfloor p \rfloor$ denotes the greatest integer less than or equal to
$p$. Such phase quantization can be implemented using $1$-bit ADCs
preceded by analog multipliers which provide linear combinations of
the $I$ and $Q$ channel samples. For instance, employing $1$-bit ADC on $I$ and $Q$ channels results in uniform $4$-sector phase quantization, while uniform $8$-sector quantization can be achieved simply by adding two new linear combinations, $I$+$Q$ and $I$-$Q$, corresponding to a $\pi/4$ rotation of $I$/$Q$ axes (no analog multipliers needed in this case), as shown in Fig. \ref{fig:Receiver}(a).\looseness-1

{\it Note:} Throughout the paper, we will assume that the PSK
constellation size $M$, and the number of quantization bins $K$, are
such that $K=aM$ for some positive integer $a$. 

   \begin{figure}[t]
      \centerline { \epsfig{figure=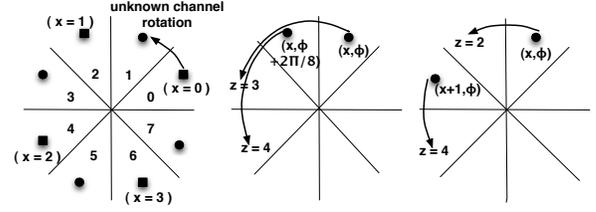, width=8cm}}
      \vspace{-3mm}
      \caption{\footnotesize QPSK with 8-sector quantization (i.e., M=4, K=8). \ a) depicts how the unknown channel phase $\phi$ results in a rotation of the transmitted symbol (square : original constellation , circle : rotated constellation). (b) and (c) depict the circular symmetry induced in the conditional probability $\prob(z|x,\phi)$ due to the circular symmetry of the complex Gaussian noise. \ \ (b) shows that increasing $\phi$ by $2\pi/K=(\pi/4)$  and $z$ by $1$ will keep the conditional probability unchanged, i.e., $\prob(z=3|x,\phi)=\prob(z=4|x,\phi+2\pi/K)$. \ \ (c) shows that increasing $x$ by $1$ and $z$ by $2 =(K/M)$ will keep the conditional probability unchanged, i.e., $\prob(z=2|x,\phi)=\prob(z=4|x+1,\phi)$.
      }
      \vspace{-3mm}
       \label{fig:Examples}
 \end{figure}

\section{Input-Output Relationship}\label{Sec:Input-Output}

In this section, we study the relationship between the channel input
and output, and present results that govern the structure of the
output probability distribution, both conditioned on the input (i.e., $\prob(\Z|\X)$), and without conditioning (i.e., $\prob(\Z)$). These distributions are integral to computing the channel capacity (our focus in this paper), as well as for soft decision decoding (not considered here).  While brute force computation (computing $\prob(\z|\x)$ for every $\z\in \cal{Z}$ and every $\x \in \cal{X}$) of these distributions has exponential complexity in the block length, we show that their inherent structure can be exploited to obtain significant complexity
reduction. We illustrate our results throughout with the running example
of QPSK with $8$-sector quantization, depicted in Fig. \ref{fig:Examples}(a). \looseness-1

Conditioned on the channel phase $\Phi$, $\prob(\Z|\X , \Phi)$ is
a product of individual symbol probabilities $\prob(Z_l|X_l,\Phi)$.  We therefore begin by analyzing the symmetries in the latter.

\subsection{Properties of $\prob(Z_l|X_l,\Phi)$}
We have that $\prob(z_l|x_l,\phi)$ is the probability that $\arg(e^{j(\theta_{x_l}+\phi)} +N_l)$ belongs to the interval $[\frac{2\pi}{K}z_l \ \ \frac{2\pi}{K}(z_l+1)).$ In other words, it is the probability that the complex Gaussian noise $N_l$ takes the point $e^{j(\theta_{x_l}+\phi)}$ on the unit circle, to another point whose phase belongs to $[\frac{2\pi}{K}z_l \ \ \frac{2\pi}{K}(z_l+1))$. Due to the circular symmetry of the complex Gaussian noise, this is the same as the probability that $N_l$ takes the point $e^{j(\theta_{x_l}+\phi+\frac{2\pi}{K}i)}$ on the unit circle, to another point whose phase belongs to $[\frac{2\pi}{K}(z_l+i) \ \ \frac{2\pi}{K}(z_l+1+i))$, where $i$ is an integer. We thus get our first two results.

{\it Property A-1:} $\prob(z_l|x_l,\phi)=\prob(z_l+i|x_l,\phi+i\frac{2\pi}{K}).$

{\it Property A-2:} $\prob(z_l|x_l,\phi)=\prob(z_l+ia|x_l+i,\phi).$

Note that $\theta_{x_l+i}=\theta_{x_l}+\frac{2\pi}{M}i= \theta_{x_l}+\frac{2\pi}{K}(ia)$, which gives Property {\it A-2}. 

Property $A$-$2$ simply states that if we jump from one point in the M-PSK constellation to the next, then we must jump $a=\frac{K}{M}$ quantization sectors in order to keep the conditional probability invariant. 
This is intuitive, since the separation between consecutive points in the input constellation is $2\pi/M$, while each quantization sector covers an angle of $2\pi/K$.
For QPSK with $K=8$, Figs. 2(b) and 2(c) depict example scenarios for the two properties.

If we put $i=-x_l$ in Property {\it A-2}, we get the following special case, which relates the conditioning on a general $x_l$ to the conditioning on $0$.

{\it Property A-3:} $\prob(z_l|x_l,\phi)=\prob(z_l-ax_l|0,\phi).$

To motivate our final property, we consider our example of QPSK with
$K=8$. While we have $8$ distinct quantization sectors, if we look at
Fig. 2(a), the orientation of these $8$
sectors relative to the $4$ constellation points (shown as squares) can be described
by dividing the sectors into $2$
groups : $\{0,2,4,6\}$, and $\{1,3,5,7\}$. For instance, the
positioning of the first sector ($z=0$) w.r.t. $x=0$ is identical to
the positioning of the third sector ($z=2$) w.r.t. $x=1$ (and
similarly $z=4$ w.r.t $x=2$, and $z=6$ w.r.t $x=3$). On the other
hand, the positioning of the second sector ($z=1$) w.r.t. $x=0$ is
identical to the positioning of the fourth sector ($z=3$) w.r.t. $x=1$
(and similarly $z=5$ w.r.t $x=2$, and $z=7$ w.r.t $x=3$). In
terms of the conditional probabilities, this implies, for example,
that we will have $\prob(z_l=7|x_l=3,\phi)=\prob(z_l=1|x_l=0,\phi)$,
and similarly, $\prob(z_l=6|x_l=3,\phi)=\prob(z_l=0|x_l=0,\phi)$.
In general, we can relate the conditional probability of every odd $z_l$ with that of $z_l=1$, and similarly of every even $z_l$ with that of $z_l=0$,
with corresponding rotations of the symbol $x_l$. For general values of $K$ and $M$, the number of groups equals $a=\frac{K}{M}$, and we can relate the probability of any $z_l$ with that of $z_l \bmod a$.\looseness-1 

{\it Property A-4:} Let $z_l= q_la+r_l$, where $q_l$ is the quotient on dividing $z_l$ by $a$, and $r_l$ is the remainder, i.e, $r_l=z_l \bmod a$. Then, $\prob(z_l|x_l,\phi)=\prob(z_l \bmod a | x_l-q_l ,\phi)$.

While this result follows directly from Property $A$-$2$ by putting
$i=-q_l$, it is an important special case, as it enables us to restrict attention to only the first $a$ sectors ($Z_l \in \{0,1,\cdots,a-1\}$), rather than having to work with all the $K$ sectors. As detailed later, this leads to significant complexity reduction in capacity computation.

We now use these properties to present results for $\prob(\Z|\X)$.
\vspace{-7mm}
\subsection{Properties of $\prob(\Z|\X)$}
\vspace{-1mm}

{\emph{Property B-1}}: Let $\pmb{1}$ denote the row vector with all entries as $1$. Then $\prob(\z|\x)=\prob(\z+i\pmb{1}|\x)$.

{\it Proof:} For a fixed $\x$, increasing each $z_l$ by the same number $i$ leaves the conditional probability unchanged, because the phase $\Phi$ in the channel model \eqref{eq:Channel_Model} is uniformly distributed in $[0,2\pi)$.
A detailed proof follows. We have
\begin{equation*}
\begin{split}
\prob(\z|\x)& =\Ex_{\Phi}\left({\prob(\z|\x,\Phi)}\right) = \Ex_{\Phi}\left({\prod_{l=0}^{L-1}{\prob(z_l|x_l,\Phi)}}\right) \\
& = \Ex_{\Phi}\left({\prod_{l=0}^{L-1}{\prob(z_l+i|x_l,\Phi+i\frac{2\pi}{K})}}\right)\\
& = \Ex_{\hat \Phi}\left({\prod_{l=0}^{L-1}{\prob(z_l+i|x_l, \hat{\Phi}}})\right)\\
& = \Ex_{\hat \Phi}\left({\prob(\z+i\pmb{1}|\x,\hat\Phi)})\right) = \prob(\z+i\pmb{1}|\x).
\end{split}
\end{equation*}
The second equality follows by the fact that the components of $\Z$ are independent conditioned on $\X$ and $\Phi$. Property $A$-$1$ gives the third equality. A change of variables, $\hat{\Phi}=\Phi+i\frac{2\pi}{K}$ gives the fourth equality (since $\Phi$ is uniformly distributed on $[0, 2\pi)$, so is $\hat{\Phi}$), thereby completing the proof. \qed  \looseness-1
{\it Remark 1:} For the rest of the paper, we refer to the operation $\z \tends\ \z+i\pmb{1}$ as {\it constant addition}.

Our next result concerns the observation that the
conditional probability remains invariant under an {\it identical}
permutation of the components of the vectors $\z$ and
$\x$. \looseness-1

{\emph{Property B-2}}: Let $\Pi$ denote a permutation operation, and $\Pi\x \ (\Pi\z)$ the vector obtained on permuting $\x \ (\z)$ under this operation. Then, $\prob(\z|\x)=\prob(\Pi{\z}|\Pi{\x})$.

{\it Proof:} As in the proof of Property $1$, the idea is to condition on $\Phi$ and work with the symbol probabilities $\prob(z_l|x_l,\Phi)$. Consider $\prob(\z|\x,\Phi) ={\prod_{l=0}^{L-1}{\prob(z_l|x_l,\Phi)}}$, and $\prob(\Pi{\z}|\Pi{\x},\Phi)= {\prod_{l=0}^{L-1}{\prob({(\Pi \z)}_l|{(\Pi \x)}_l,\Phi)}}$. Since multiplication is a commutative operation, we have $\prob(\z|\x,\Phi)=\prob(\Pi{\z}|\Pi{\x},\Phi)$. Taking expectation w.r.t. $\Phi$ completes the proof. \qed

The next two results extend properties $A$-$3$ and $A$-$4$.

{\it Property B-3:} Define the input vector $\x_0=[0 \cdots 0]$. Then, $\prob(\z|\x)=\prob(\z-a\x|\x_0)$, where $a=\frac{K}{M}$, and the subtraction is performed modulo $K$.

{\it Property B-4:} Let $z_l= q_la+r_l$, where $q_l$ is the quotient on dividing $z_l$ by $a$, and $r_l$ is the remainder, i.e, $r_l=z_l \bmod a$. Define $\q=[q_0,\cdots,q_{L-1}]$, and, $\z \bmod a =[z_0 \bmod a \ \  \cdots \ \ z_{L-1} \bmod a]$. Then $\prob(\z|\x)=\prob(\z \bmod  a \ | \ \x-\q) $.

{\it Proofs:} The properties follow from $A$-$3$ and $A$-$4$ respectively, by first noting that the vector probability $\prob(\z|\x,\Phi)$ is the product of the scalar ones, and then integrating over $\Phi$ .\qed
\vspace{-1mm}
\subsection{Properties of $\prob(\Z)$}
\vspace{-1mm}

We now consider the unconditional distribution $\prob(\z)$. The first
result states that $\prob(\z)$ is invariant under constant addition.

{\it Property C-1:} $\prob(\z)=\prob(\z+i\pmb{1})$.

{\it Proof:} Using Property $B$-$1$, this follows directly by taking expectation over $\X$ on both sides.
\qed

On similar lines, we now extend Property $B$-$2$ to show that $\prob(\z)$ is invariant under any permutation of $\z$.

{\emph{Property C-2:}} $\prob(\z)=\prob(\Pi{\z})$.

{\it Proof:} We have $\prob(\z) =\frac{1}{M^L}\sum_{\x \in \mathcal{X}}{\prob(\z|\x)}$. Using Property $B$-$2$, we get $\prob(\z) =\frac{1}{M^L}\sum_{\x \in \mathcal{X}}{\prob\left(\Pi{\z}|\Pi\x\right)}.$ Since $\Pi$ is just a permutation operation, every unique choice of $\x \in \mathcal{X}$ results in a unique $\Pi\x \in \mathcal{X}$. Hence, we can rewrite the last equation as $\prob(\z) \ = \ \frac{1}{M^L}\sum_{\x \in \mathcal{X}}{\prob(\Pi{\z}|\x}) \ = \ \prob(\Pi{\z}).$ \qed

Our final result extends Property $B$-4.

{\it Property C-3}:  Let $a =\frac{K}{M}$. Then $\prob(\z)=\prob(\z \bmod a)$.

{\it Proof:} Using the same notation as in Property $B$-$4$, we have $\prob(\z|\x)=\prob(\z \bmod  a \ | \ \x-\q) \ $. Noting that the transformation $\x \tends \x-\q$ is a one-to-one mapping, the proof follows on the same lines as the proof of Property $C$-$2$.\qed

{\it Example:} For QPSK with $K=8$ and $L=4$, $\prob(z=[5 \ 7 \ 2 \ 4])=\prob(z=[1 \ 1 \ 0 \ 0])$.

We now apply these results for low complexity capacity computation.
\vspace{-1mm}

\section{Efficient Capacity Computation}\label{Sec:Capacity}

We wish to compute the mutual information 
\[
I(\X;\Z)= H(\Z)-H(\Z|\X) .
\]
We first discuss computation of the conditional entropy.

\vspace{1.2mm}
\noindent{\it A . \ Computation of the conditional entropy $H(\Z|\X)$}
\vspace{0.6mm}

We have $H(\Z|\X)=\sum_{\cal{X}} H(\Z|\x)\prob(\x)$, where
$H(\Z|\x)=-\sum_{\cal{Z}} \prob(\z|\x)\log\prob(\z|\x)$ is the entropy
of the output when the input vector $\X$ takes on the specific value
$\x$. Our main result in this section is that $H(\Z|\x)$ is constant
$\forall \x$.

{\emph{Property D-1}}: $H(\Z|\x)$ is a constant.

{\it Proof:} We show that for any input vector $\x, H(\Z|\x)=H(\Z|\x_0)$, where $\x_0=[0 \cdots 0]$ as defined before. We have
\begin{eqnarray}
H(\Z|\x) =&-\displaystyle&\sum_{\cal{Z}} \prob(\z|\x)\log\prob(\z|\x) \nonumber \\
=&-\displaystyle&\sum_{\cal{Z}} \prob(\z-a\x|\x_0)\log\prob(\z-a\x|\x_0) \label{eq:Cond_Entropy} \ ,
\end{eqnarray}
where the second equality follows from Property $B$-$3$. Now, since $\z \tends \z-a\x$ is just a subtraction operation, it is easy to see that every unique choice of $\z \in \cal{Z}$ results in a unique choice of $\z-a\x \in \cal{Z}$. Hence, we can rewrite \eqref{eq:Cond_Entropy} as
\begin{equation}
\begin{split}
H(\Z|\x)& =-\displaystyle\sum_{\cal{Z}} \prob(\z|\x_0)\log\prob(\z|\x_0) = H(\Z|\x_0)
\end{split}
\end{equation}
\qed

Thus, $H(\Z|\X) = H(\Z|\x_0)$, but brute force computation of
$H(\Z|\x_0)$ still has exponential complexity, $\prob(\Z|\x_0)$ must
be computed for each of the $K^L$ possible output vectors
$\Z$. However, we show that it suffices to compute $\prob(\Z|\x_0)$ for a much
smaller set of $\Z$ vectors. \looseness-1

Using Property $B$-$2$, we have $\prob(\z|\x_0)=\prob(\Pi \z|\Pi \x_0)$. Since $\x_0=[0 . . 0]$, any permutation of $\x_0$ gives back $\x_0$. Hence, $\prob(\z|\x_0)=\prob(\Pi{\z}|\x_0)$. Combined with Property $B$-$1$, we thus get that it suffices to compute $\prob(\z|\x_0)$ for a set of vectors $S_{\Z}$ in which no vector can be obtained from another by performing the operations of constant addition and permutation. For $K=8$ and $L=\{3,4,5,6,7\}$, the cardinality of the entire set of $\Z$ vectors, $K^{L}$, evaluates to $\{512, 4096, 32768, 2.6\times 10^5, 2.1\times 10^6\}$, while the cardinality of $S_\Z$ is $\{15,43,99,217,429\}$, illustrating the large reduction in complexity.
For simplicity of exposition, we do not delve into the exact details of how we can obtain the set $S_\Z$. Fast algorithms to do this, and their associated complexity are currently being investigated. More details are available from the authors upon request, and will be provided in future publications as well. \looseness-1

Once we have the set $S_\Z$, we can numerically compute the probability $\prob(\z|\x_0)$ for every vector in $S_\Z$. The entropy $H(\Z|\x_0)$ can then be obtained as follows. For $\z \in S_\Z$, let $n(\z)$ denote the number of distinct vectors that can be generated from it by performing the operations of constant addition and permutation. This is straightforward to compute. The conditional entropy then is $H(\Z|\x_0)=
-\displaystyle\sum_{S_\Z}n(\z) \prob(\z|\x_0)\log\prob(\z|\x_0)$. \\

\noindent{\it B. \ Computation of the output entropy $H(\Z)$}

The output entropy is $H(\Z)=-\sum_{\cal{Z}} \prob(\z)\log\prob(\z)$. A brute force computation requires us to know $\prob(\z) \ \forall \z \in \cal{Z}$, which clearly has exponential complexity. However, using Properties $C$-$1$, $C$-$2$ and $C$-$3$, we get that it is sufficient to compute $\prob({\z})$ for a set of vectors $\tilde{S}_\Z$ in which no vector can be obtained from another one by performing the operations of constant addition and permutation, and also, the vector components $\in \{0,1,\cdots,a-1\}$.  This is similar to the situation we encountered earlier in the last subsection, except that the vector components there were allowed to be in $\{0,1,\cdots,K-1\}$. 

{\it Example:} For QPSK with $8$ sectors (so $a=2$), the relevant vectors for block length $2$ are $[0 \ \ 0]$ and $[0 \ \ 1]$.

\subsubsection*{Computation of $\prob({\Z})$}
For each of the vectors in the set $\tilde{S}_\Z$ defined above, we now need to obtain $\prob(\z)=\sum_{\x \in \cal{X}}\prob(\z|\x)\prob(\x)$. A brute force approach is to compute $\prob(\z|\x)$ for each $\x$. However, we can exploit the structure in $\z$ to reduce the number of vectors $\x$ for which we need $\prob(\z|\x)$. Specifically, we have that 
each $z_i \in \{0,1,\cdots,a-1\}$. Since there are only $a$ different types of components in $\z$, for block length $L>a$, some of the components in $\z$ will be repeated. For any $\x$, we can then use Property $B$-$2$ to rearrange the components at those locations for which the components in $\z$ are identical, without changing the conditional probability. For instance, let $z_m=z_n$ for some $m,n$. Then, $\prob(\z|\x)=\prob(\z|\Pi{\x})$, where $\Pi{\x}$ is obtained from $\x$ by rearranging the components at locations $m$ and $n$. To sum up, we can restrict attention to a set of vectors $S_\X$ in which no vector can be obtained from another one by permutations between those locations for which the elements in $\z$ are identical. While for large $a$, the potential reduction in complexity may not be large, for small values of $a$ (which is the paradigm of interest in this work), the savings will be significant. As before, the algorithmic details for obtaining the set $S_\X$ will be provided in upcoming publications.

\vspace{-1mm}
\section{Numerical Results}
We now present capacity results (obtained using the low-complexity procedure outlined in the last section) for QPSK with 8-sector and 12-sector phase quantization, for different block lengths L. For all our results, we normalized the mutual information $I(\X;\Z)$ by $L$-1 to obtain the per symbol capacity, since in practice the successive blocks can be overlapped by one symbol due to slow phase variation from one block to the next.\looseness-1

\begin{figure}[]
      \centerline { \epsfig{figure=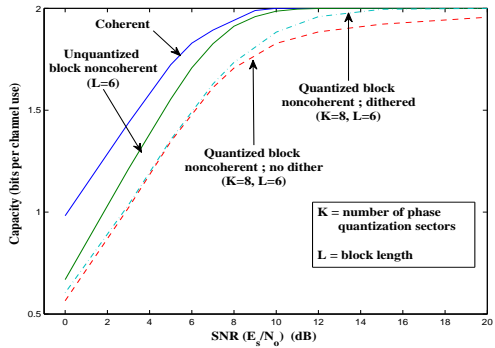, width=7.6cm, height=5cm}}
      \vspace{-3mm}
      \caption{\footnotesize Performance comparison for QPSK with block length $L=6$ : plots depict the capacity of the block noncoherent channel without quantization, and with 8-sector quantization (with and without dithering). Also shown is the capacity for coherent QPSK.}
       \label{fig:Results_8_sector}
       \vspace{-5.3mm}
\end{figure}

{\it 8-sector quantization:} In Fig. \ref{fig:Results_8_sector}, we plot the channel capacity with 8-sector quantization, at different SNR values .(To avoid clutter, we show the results for $L=6$ only.) Also shown for reference are the capacity values for the coherent case, and for the block noncoherent case without any quantization. We see that, for $\snr >$ 2-3 dB, our simple $8$-sector quantization scheme recovers more than $80$-$85$\% of the spectral efficiency obtained with unquantized observations. This is encouraging, given that our work is targeted towards future high bandwidth systems (such as those operating in the $60$ GHz mm-wave band), for which a small reduction in spectral efficiency is acceptable. On the other hand, if we measure the power loss for fixed spectral efficiency, we see that at rates of up to about $1.2$ bits/channel use, there is a loss of about $1$-$1.5$ dB compared to the unquantized case. However, the loss is more significant as $\snr$ increases: the capacity approaches $2$ bits/channel use rather slowly at high $\snr$. Since the input entropy $H(\X)$ is constant, this in turn implies that $H(\X|\Z)$ falls off very slowly as $\snr \tends \infty$. A more detailed analysis of the likelihood ratio $\prob(\Z|\X,\Phi)$ (omitted here due to lack of space) provides insight into this behavior. We find that in addition to the symmetries in $\prob(\Z|\X,\Phi)$ that we exploited to reduce the complexity of capacity computations, there are certain other symmetries with adverse consequences as well : they make it impossible to distinguish between the effect of the unknown phase offset and the phase modulation on the received signal. More specifically, if we consider the maximum likelihood (ML) estimator $\displaystyle \argmax_{\x, \phi}\ \prob(\z|\x,\phi)$, we find that for certain outputs $\z$, irrespective of the $\snr$ (and also the block length), the estimator always returns two distinct equally likely solutions $(\x_1,\phi_1)$ and $(\x_2,\phi_2)$. In an information-theoretic sense, this ambiguity indicates a significant conditional entropy $H(\X|\Z)$. As $\snr \tends \infty$, the probability of these ambiguous outputs does go to zero, but very slowly, leading to a slow decrease in $H(\X|\Z)$ as well.\looseness-1

\begin{figure}[t]
      \centerline { \epsfig{figure=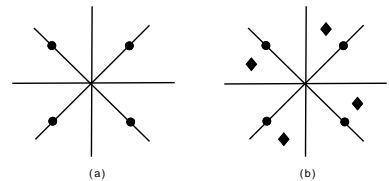, width=5cm}}
      \vspace{-3mm}
      \caption{\footnotesize (a) Standard PSK : the same constellation (the one shown) is used for both symbols in the block. (b) Dithered-PSK : the constellations used for the two symbols are not identical, but the second one is a dithered version of the first one.}
       \label{fig:Dither}
       \vspace{-3mm}
\end{figure}

Possible ways to break the undesirable symmetries could be to use non-uniform phase quantization, or to employ dithering across symbols in a block. Here we investigate the role of the latter. We can dither either at the transmitter by rotating the QPSK constellation points, or at the receiver by using analog pre-multipliers to shift the phase quantization boundaries. We use a simple transmit dither scheme in which we rotate the QPSK constellation by an angle of $\frac{1}{L}(\frac{2\pi}{K})$ from one symbol to the next. Fig. \ref{fig:Dither}(a) shows this scheme for block length L=2 and K=8. The constellation used for the second symbol (shown by the diamond shape) is dithered from the constellation used for the first symbol by an angle of $\pi/8.$ With this choice of transmit constellations, we find that the ambiguity in the ML estimator is removed, and hence the performance is expected to improve. The plot in Fig. \ref{fig:Results_8_sector} shows the performance improvement for L=6. {\footnote {Since the low-complexity procedure outlined in Section \ref{Sec:Capacity} does not work once we dither, we used Monte Carlo simulations to compute the capacity with dithering.}}

While the preceding simple transmit dither scheme has improved the performance for 8-sector quantization, we hasten to add that there is no optimality associated with it. A more detailed investigation of different dithering schemes and their potential gains is therefore an important topic for future research.\looseness-1

\begin{figure}[]
      \centerline { \epsfig{figure=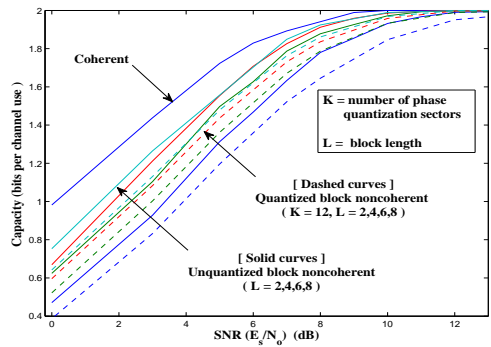, width=7.6cm, height=5cm}}
      \vspace{-3mm}
      \caption{\footnotesize Performance comparison for QPSK : plots depict the capacity of the coherent channel, unquantized block noncoherent channel (different block lengths), and the 12-sector quantized block noncoherent channel (different block lengths).}
       \label{fig:Results_12_sector}
       \vspace{-5.5mm}
\end{figure}

{\it 12-sector quantization:} In Fig. \ref{fig:Results_12_sector}, we plot the performance curves for QPSK with 12-sector quantization, for block length L=2,4,6,8. Also shown for reference are the coherent and unquantized block noncoherent performance curves. For identical block lengths, the loss in capacity (at a fixed $\snr >$ 2-3 dB) compared to the unquantized case is less than 5-10 \%, while the loss in power efficiency (for fixed capacity) varies between 0.5-2 dB, and dithering is not required.\looseness-1

\vspace{-1mm}
\section{Conclusions}
\vspace{-1mm}
We have investigated the capacity limits imposed by the use of low-precision phase quantization at a carrier-asynchronous receiver. The symmetries in input-output relationship of the resulting channel have been exploited to reduce the complexity of capacity computation. Important topics for future research include a more detailed investigation of different dithering schemes (motivated by the performance improvement obtained using the simple scheme considered here), as well as development of practical capacity-approaching coded modulation strategies. An important practical issue is determining whether timing synchronization (which is assumed in the model here) can also be attained using phase-quantized samples, or whether some form of additional information (perhaps using analog techniques prior to the ADC) is required.

\bibliographystyle{IEEEbib}
\bibliography{strings,refs,manuals}

\appendices


\end{document}